%% file: PaperICRC2023.tex
\documentclass[a4paper,11pt]{article}
\usepackage[numbers,square,sort&compress]{natbib}
\usepackage{pos}
\usepackage{hyperref}
\usepackage{listings}
\usepackage{lineno}
\usepackage{subcaption}
\usepackage{graphicx}
\usepackage{enumitem}

\newcommand{\cobj}{\lstinline }

\title{Parallel processing of radio signals and detector arrays in CORSIKA 8}
 \ShortTitle{Parallel processing of radio signals and detector arrays in CORSIKA 8}

\author*[a]{A. Augusto Alves Jr}
\author[a]{Nikolaos Karastathis}
\author[a,b]{Tim Huege}

\affiliation[a]{Institute for Astroparticle Physics (IAP), Karlsruhe Institute of Technology, Karlsruhe, Germany}
\affiliation[b]{Astrophysical Institute, Vrije Universiteit Brussel, Belgium}

\onbehalf{for the CORSIKA 8 collaboration} 

\emailAdd{aalvesju@gmail.com}

\abstract{This contribution describes some recent advances in the parallelization of the generation and processing of radio signals emitted by particle showers in CORSIKA 8. CORSIKA 8 is a Monte Carlo simulation framework for modeling ultra-high energy particle cascades in astroparticle physics. The aspects associated with the generation and processing of radio signals in antennas arrays are reviewed, focusing on the key design opportunities and constraints for deployment of multiple threads on such calculations. The audience is also introduced to Gyges, a lightweight, header-only and flexible multithread self-adaptive scheduler written compliant with C++17 and C++20, which is used to distribute and manage the worker computer threads during the parallel calculations. Finally, performance and scalability measurements are provided and the integration into CORSIKA 8 is commented.}

\ConferenceLogo{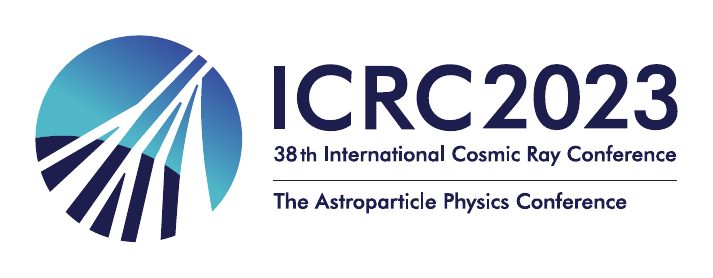}

\FullConference{%
38th International Cosmic Ray Conference (ICRC2023)\\
  26 July - 3 August, 2023\\
  Nagoya, Japan}

\begin{document}
\maketitle

\section{Introduction}
\label{introduction}

Over the past couple of decades of research on extensive particle showers, radio detection has become a technique competitive with
standard particle and fluorescence driven measurements. Due to the complexity
of extensive particle showers, in air and other media, detailed particle-level simulations of the radio emissions
are often needed to analyze experimental data and reconstruct the properties of the primary particles.

In this context, the two standard software tools used for radio emission simulations are
CoREAS \cite{Simulating-radio-emission-from-air-showers-with-CoREAS} as implemented in CORSIKA 7 and ZHAireS \cite{PhysRevD.81.123009}. These tools implement two
different formalisms for calculating the radio emission from the particle tracks in the extensive
particle shower, namely the “Endpoint” \cite{PhysRevE.84.056602,LUDWIG2011438} and the “ZHS” \cite{ PhysRevD.45.362} formalisms, respectively. Both algorithms
have been recently implemented on CORSIKA 8 \cite{Engel_2018}, which is a modern C++17 compiliant Monte Carlo simulation framework for modeling
ultra-high energy particle cascades in astroparticle physics. 

Additionally, proposed next-generation experiments with growing array size and channel-count pose significant challenges
regarding the computational cost for calculating radio emissions, especially for ultra high-energy showers and signals propagating in media
with varying properties. In order to mitigate such impacts, the radio emission
module for the CORSIKA 8 (C8) framework \cite{icrc21} has been reimplemented in multithread friendly fashion. This contribution discusses
these developments and is organized as the following. \autoref{sec:radio-overview} gives an overview of the radio module of CORSIKA 8.
\autoref{sec:gyges}, Gyges, a C++17/20 library for distribution and management of tasks on multithread systems, is presented. 
\autoref{sec:parallelization} the parallelization strategy used of the radio module calculations is covered, including the corresponding updates 
on the interfaces and codes implementing the algorithms. Finally, \autoref{sec:performance} presents the performance gains in function of the number of threads for both formalisms, measured for array detectors with different sizes. \autoref{sec:conclusions}, the conclusions and perspectives are drawn.

\section{Overview of the radio module in CORSIKA 8}
\label{sec:radio-overview}

\begin{figure}[tb!]
 \centering
 \includegraphics[width=0.85\linewidth]{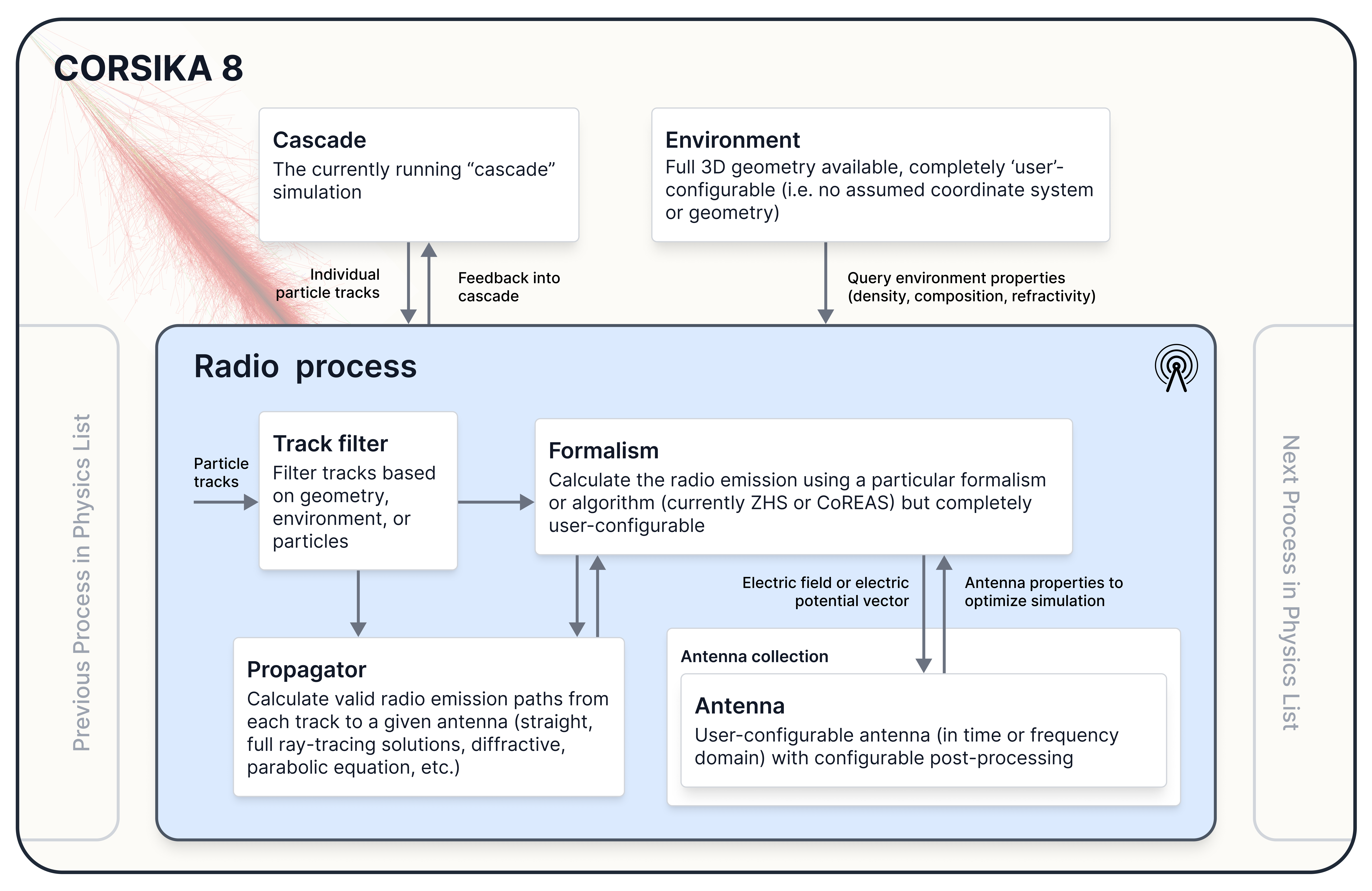}
 \caption{A schematic diagram of the radio process currently implemented in CORSIKA 8 and how it integrates with the CORSIKA 8 framework} 
 \label{fig:radio_module}
\end{figure}

The top-level architecture of the radio process module is shown in  \autoref{fig:radio_module}. 
All components in the module can be independently configured and combined with either the CORSIKA 8 built-in interface or custom C++ code, making possible construct multiple radio process instances for different scenarios. The components of the modules have been extensively presented in \cite{icrc21,arena22}.
In this contribution, the flow of the radio calculation is discussed and how performance is enhanced using multithreading. 

Once the radio process has received a particle track, the track is being checked according to the \textit{track filter} in order to be determined if this track is relevant for the radio calculation or not. The track is then pushed forward to the \textit{formalism} and the track needs to be looped over all antennas existing in the antenna collection. A significant portion of the calculation happens after this step, which needs to be repeated for every antenna available and for every single particle track provided. This is precisely the part of the code we wish to accelerate with this work. Inside the loop, the particle track is fed to the propagator, which calculates the valid emission paths from the particle to the antenna. Hence, all the necessary information to calculate the electric field vector (or vector potential) is present now, and finally this information is processed and stored in the \textit{antenna} instance. The load of this calculation is directly affected by the underlying complexity of the \textit{propagator} used. Naturally, the larger the number of antennas in the detector, the higher the runtime of the radio simulation will be. By assigning different bunches of antennas to available threads, we expect to observe a significant performance boost.

\section{Gyges}
\label{sec:gyges}

Gyges is a lightweight C++17, or higher, header-only library to manage thread pooling, which has been developed in the context of the ongoing effort to paralellize the CORSIKA 8 framework.
By deploying Gyges, the computational costs associated to creating and destroying a thread-pool, a \cobj{gyges::gang} in the library's jargon, can be paid just once in the program lifetime, with threads of the pool picking-up tasks as they become available. If there are no tasks, the threads just go sleeping. Additionally, tasks can be submitted from multiple threads, with the submitter getting a \cobj{std::future} object to monitor the task in-place. On the task implementation side, developers get access to a \cobj{std::stop_token} that can be used to interrupt the task execution, if a request to do so arrives from \cobj{gyges::gang} via the \cobj{gyges::gang::stop()}.

As default behavior, once a \cobj{gyges::gang} is created, it will promptly pick up and process any submitted task. This behavior can be changed, putting the \cobj{gyges::gang} in a “hold-on” state. In that case, the processing of the tasks will
be postponed until it is put back on “unhold” status, while the threads will be put to sleep until the \cobj{gyges::gang::unhold()} command is sent. Among other features, Gyges provides two implementations of the \cobj{gyges::for_each} algorithm, with one of than able to use an already existing \cobj{gyges::gang} object.

Gyges is licensed under GPL version 3 and is currently in a stable release state. The code is available at \url{https://gitlab.iap.kit.edu/AAAlvesJr/Gyges}.

\begin{lstlisting}[language=c++, caption={Interface of gyges::gang and gyges::for\_each implementations.}, label={lst:gyges}, captionpos=b, frame=shadowbox,float=t]

class gang
{
    // constructor taking the 
    gang(unsigned int const thread_count=
           std::thread::hardware_concurrency(),
    bool release = true ) ; 
    gang( gang const & other ) = delete;
    gang( gang && other )      = delete;
    //submit a task implementing void operator( void )
    template<typename FunctionType>
    inline std::future<void> submit_task(FunctionType f) ;
    //notify the running tasks (request stop), 
    inline void stop(void);
    //put the gang on ``hold'' status
    inline void hold(void);
    //revert the gang to ``processing'' status
    inline void unhold(void);
    //checks the gang status
    inline bool on_hold(void);
    //get the gang size
    inline std::size_t size(void);
};

// for_each accepting a pre-created gang
template<typename Iterator, typename Predicate>
void for_each(Iterator begin, Iterator end,
           Predicate const& functor,  gang& pool);
// for_each
template<typename Iterator, typename Predicate>
void for_each(Iterator begin, Iterator end, 
           Predicate const&  functor);
\end{lstlisting}

\section{Radio module parallelization strategy}
\label{sec:parallelization}

The radio module calculates the signal corresponding to each particle, and the tracks that describe its trajectory,
for each antenna of the array detector, often running as one of the final operations in the particle simulation process sequence.
In order to parallelize the radio module, the calculation of the signal over the array detector is processed using a \cobj{gyges::gang}
containing a specifiable number of threads, in a such way that, for each particle and its tracks, the response of the antennas and the storing of information is 
calculated in parallel. 

Since the signal processing corresponding to a single antenna is not intensive enough to occupy efficiently a thread,
each submitted task computes the response corresponding to a bunch of antennas. As it will be detailed in
\autoref{sec:performance}, the number of antennas in this bunch in comparison to the Gyges gang size is a critical parameter for the overall efficiency of the radio module. 

This logic is implemented with the introduction of a couple of classes, one per formalism, to encapsulate the pulse calculated for each antenna in a callable object abstracting away the implementation details
of CoREAS and ZHAireS. This object is called runner, and it is the one to be distributed, together with the antenna collection that describes the array detector, to the worker threads managed by the \cobj{gyges::gang} instance, which is being held by the \cobj{corsika::RadioProcess} and has the same life-time of it. These developments are complemented by changes in the user interfaces easing to instate and to deploy the radio module. These changes are summarized in \autoref{lst:interface}.

CORSIKA 8-wise, the expected overall speed-up depends hugely on the detector size, i.e. number of antennas in the detector. For large array detectors, or computing intensive propagators, the importance of radio module operations grows, tending to dominate the particle simulation sequence. In such situations, the speed-up is larger.

\begin{lstlisting}[language=c++, caption={Improved interface of radio module.}, label={lst:interface}, captionpos=b, frame=shadowbox,float=t]%\begin{minted}[mathescape, numbersep=5pt,frame=lines, framesep=2.5mm]{c++}

//convenience function for creating a propagator,
//taking as parameter an environment object 
auto propagator = make_simple_radio_propagator(environment);
//convenience functions for creating CoREAS and ZHS instances,
//taking as parameters detector
//and propagator objects, as well as the number of threads
auto coreas = make_radio_process_CoREAS(detector, propag, nthreads);
auto    zhs = make_radio_process_ZHS(detector, propag, nthreads);
\end{lstlisting}

\section{Performance measurements and validation}
\label{sec:performance}

\begin{figure}[tb!]
\centering
\begin{subfigure}{0.49\textwidth}
    \includegraphics[width=\textwidth]{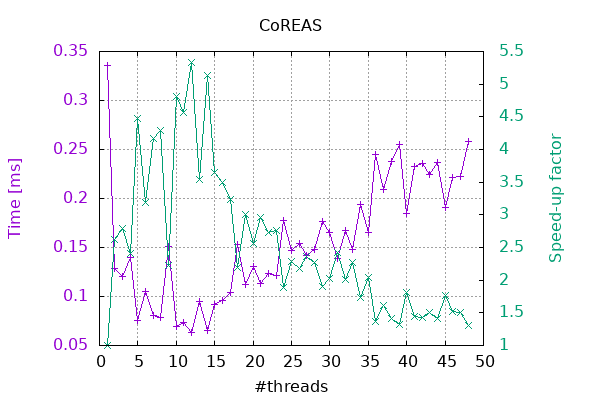}
    \label{fig:performance_per_particle_1}
\end{subfigure}
\hfill
\begin{subfigure}{0.49\textwidth}
    \includegraphics[width=\textwidth]{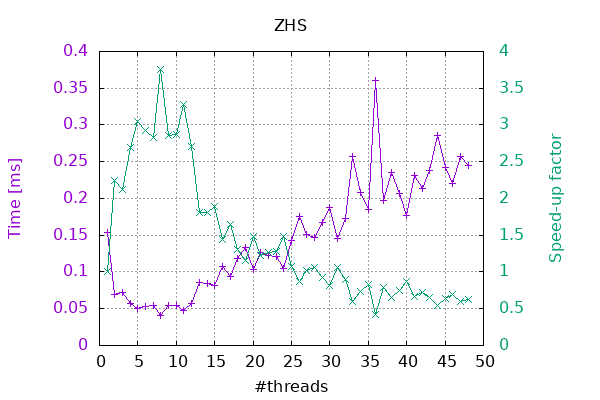}
    \label{fig:performance_per_particle_2}
\end{subfigure}
\caption{Performance to process a single particle as a function of number of threads for an array detector containing 200 antennas.}
\label{fig:performance_per_particle_200}
\end{figure}

\begin{figure}[tb!]
\centering
\begin{subfigure}{0.49\textwidth}
    \includegraphics[width=\textwidth]{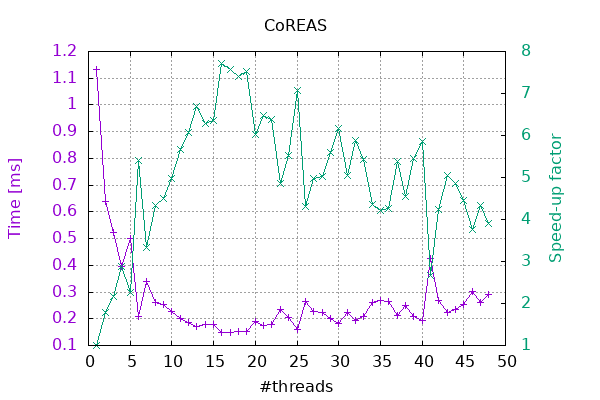}
    \label{fig:performance_per_particle_3}
\end{subfigure}
\hfill
\begin{subfigure}{0.49\textwidth}
    \includegraphics[width=\textwidth]{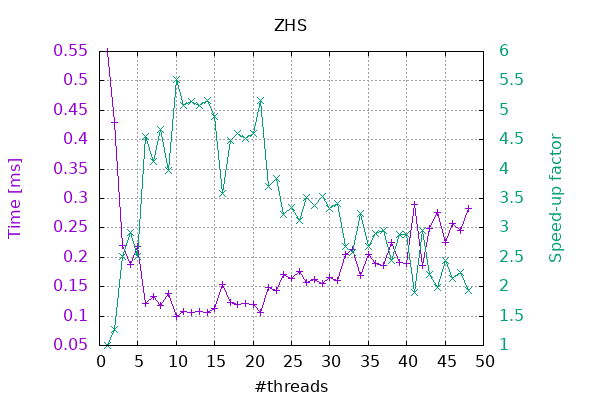}
    \label{fig:performance_per_particle_4}
\end{subfigure}
\caption{Performance to process a single particle as a function of number of threads for an array detector containing 1000 antennas.}
\label{fig:performance_per_particle_1000}
\end{figure}

\begin{figure}[tb!]
\centering
\begin{subfigure}{0.49\textwidth}
    \includegraphics[width=\textwidth]{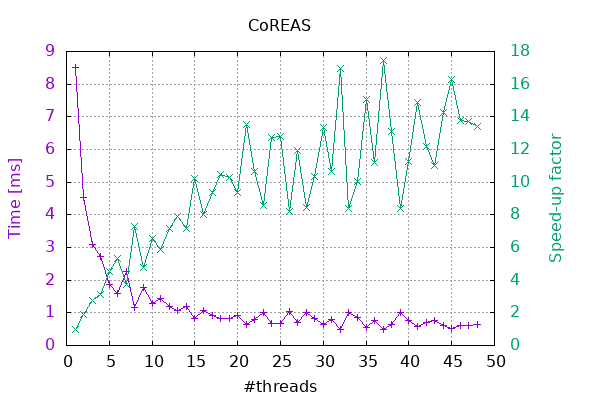}
    \label{fig:performance_per_particle_5}
\end{subfigure}
\hfill
\begin{subfigure}{0.49\textwidth}
    \includegraphics[width=\textwidth]{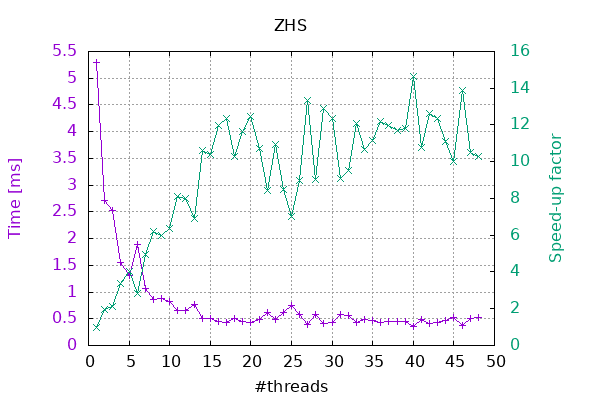}
    \label{fig:performance_per_particle_6}
\end{subfigure}
\caption{Performance to process a single particle as a function of number of threads for an array detector containing 10,000 antennas.}
\label{fig:performance_per_particle_10000}
\end{figure}

The raw performance gains from parallelization of the radio module calculations over the antennas of the array detector have been assessed measuring the time spent, and the corresponding speed-up, to process the electromagnetic pulse from a single particle as a function of the array detector size and number of threads. 
Array detectors with different sizes have been tested against \cobj{gyges::gang} with up to 48 worker threads.
The results are summarized in \autoref{fig:performance_per_particle_200}, \autoref{fig:performance_per_particle_1000}  and  \autoref{fig:performance_per_particle_10000}

\autoref{fig:performance_per_particle_200} shows that for array detectors with 200 antennas, the speed-up peaks between 10 and 15 worker threads, beyond which the performance decreases due to computing tasks not being able to occupy the CPU enough to hide the latency and costs associated to management of multiple threads. As it is shown in \autoref{fig:performance_per_particle_1000} and \autoref{fig:performance_per_particle_10000}, by increasing the number of antennas, the speed-up scales mostly as predicted by Amdahl's law. Similar results would be achieved, albeit leading to performance peaking at different number of threads, when deploying propagators performing heavier calculations. 

\begin{figure}[tb!]
 \centering
 \includegraphics[width=0.6\textwidth]{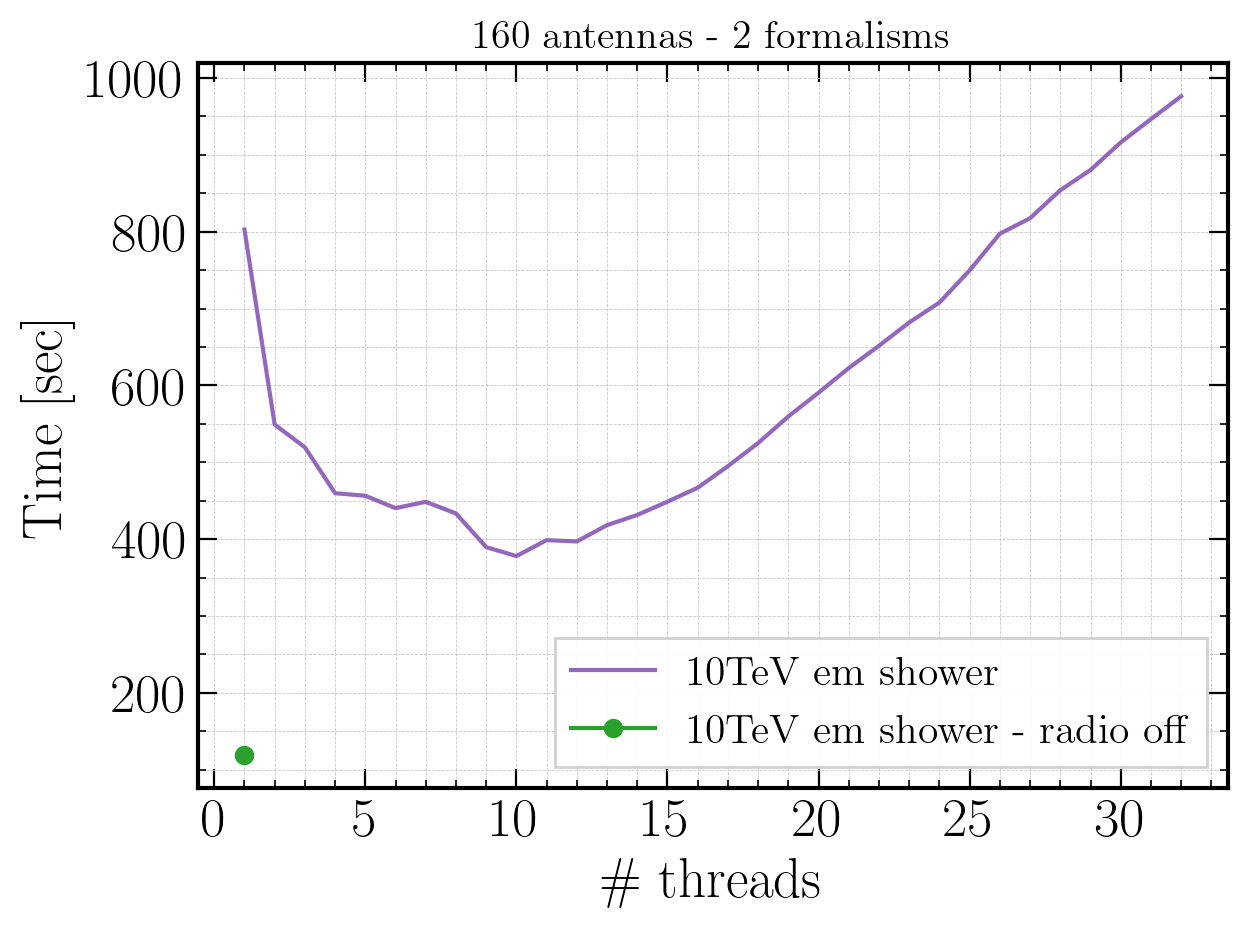}
 \caption{The parallelized radio module running on an electron induced air shower processing a detector array of 160 antennas. 2 formalism, namely CoREAS and ZHS are activated and use 160 antennas each. The performance peaks at 10 worker threads, beyond which the performance degradates.}
  \label{fig:full_shower}
\end{figure}

The overall impact of the parallelization of the radio module on CORSIKA 8 has been measured running a full electromagnetic shower simulation. In that scenario, due to the Gyges design, the overhead for creating, managing and submitting tasks to the thread pool is negligible in comparison to the other initialization routines called up-front in the full shower simulation. The radio module is currently the only component of the CORSIKA 8 sequence capable of performing its tasks in parallel, meaning that the maximum speed-up is limited by the amount of code running sequentially, in accordance with Amdhal's law. The total time to run the full shower is limited below by not deploying the radio module at all, and above by running this module in a single thread, that is sequentially. \autoref{fig:full_shower} summarizes the results and confirms
the profiles performed for measuring the single particle performance. In the same figure, we show for reference the runtime of the same electron induced shower with the radio emission calculation turned off.

\begin{figure}[tb!]
\centering
\begin{subfigure}{4.9cm}
    \includegraphics[width=4.9cm]{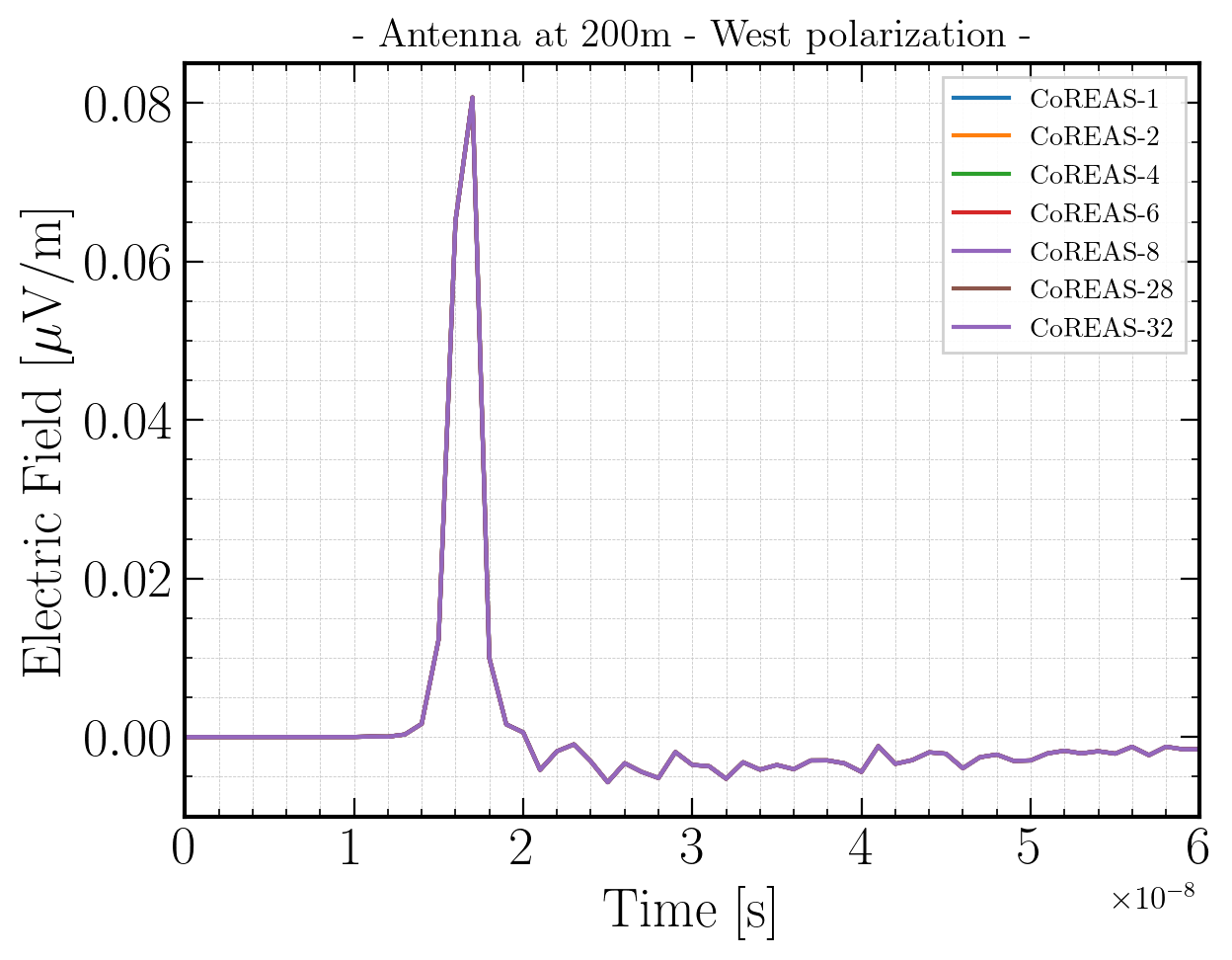}
    \label{fig:west-coreas}
\end{subfigure}
\begin{subfigure}{4.9cm}
    \includegraphics[width=4.9cm]{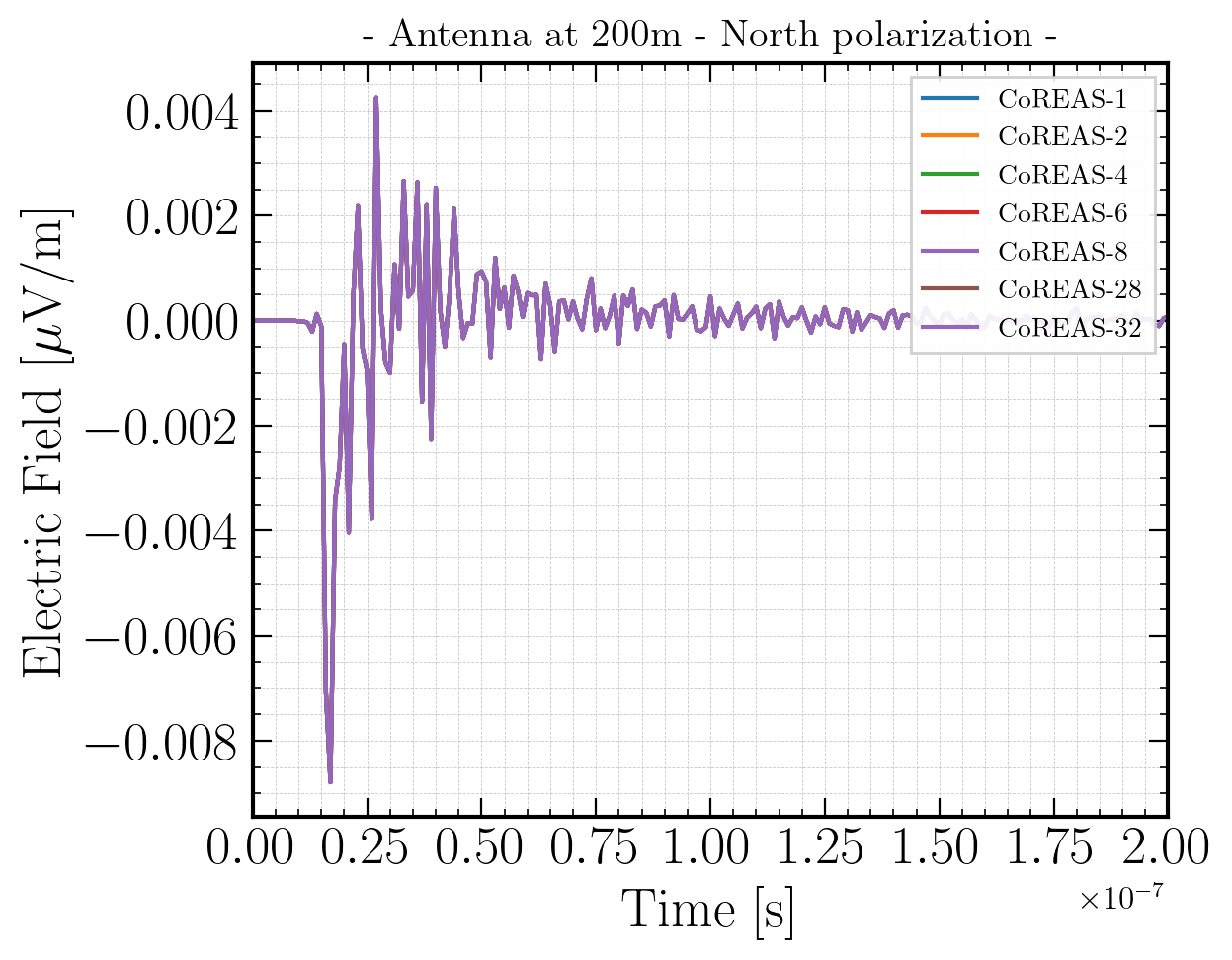}
    \label{fig:north-coreas}
\end{subfigure}
\begin{subfigure}{4.9cm}
    \includegraphics[width=4.9cm]{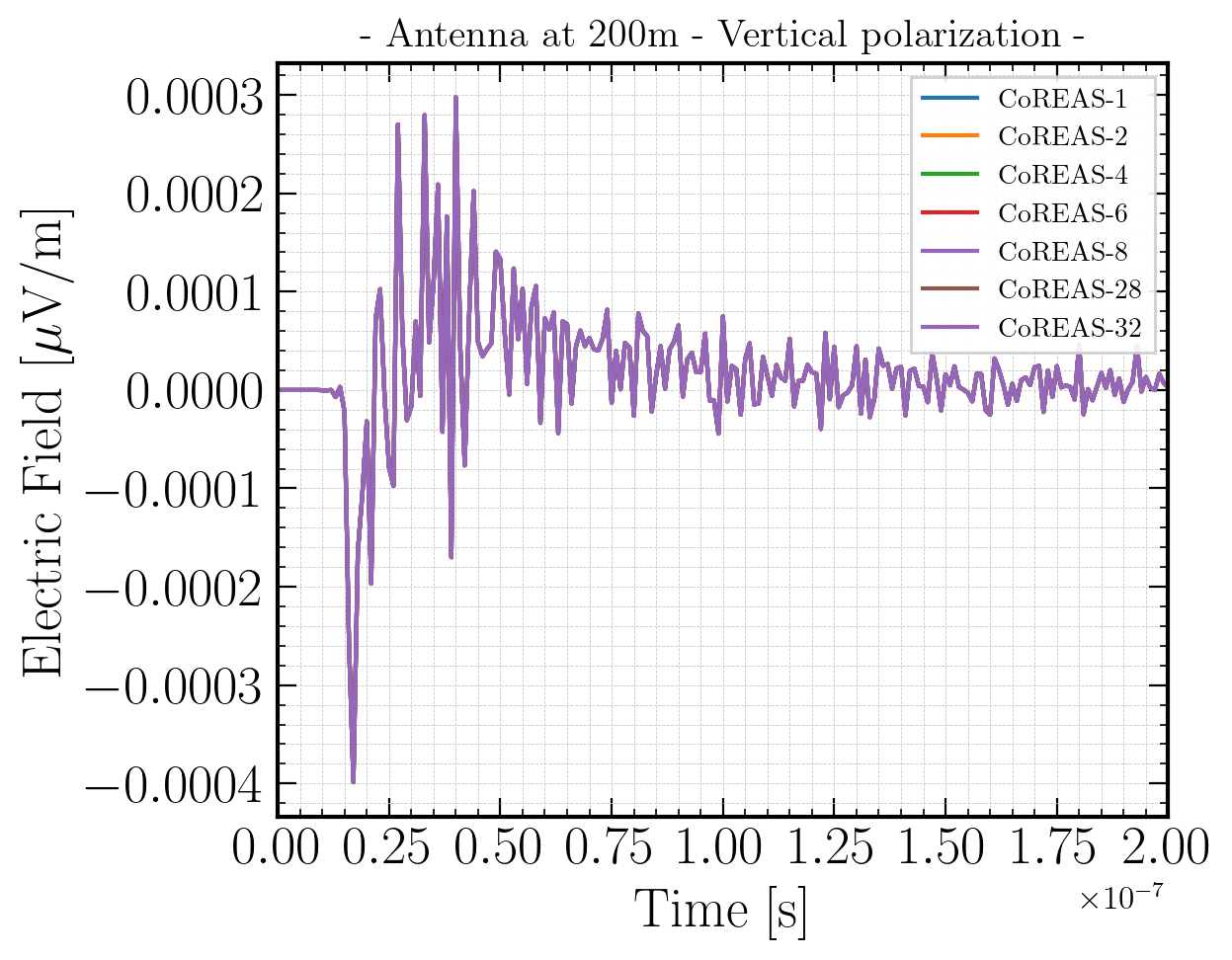}
    \label{fig:vertical-coreas}
\end{subfigure}
\caption{Pulse comparisons in all three polarizations using the CoREAS formalism. The pulses have been simulated and processed on \cobj{gyges::gangs} of different sizes. Different number of threads produce identical pulses for CoREAS, as expected.}
\label{fig:polarizations-coreas}
\end{figure}

\begin{figure}[tb!]
\centering
\begin{subfigure}{4.9cm}
    \includegraphics[width=4.9cm]{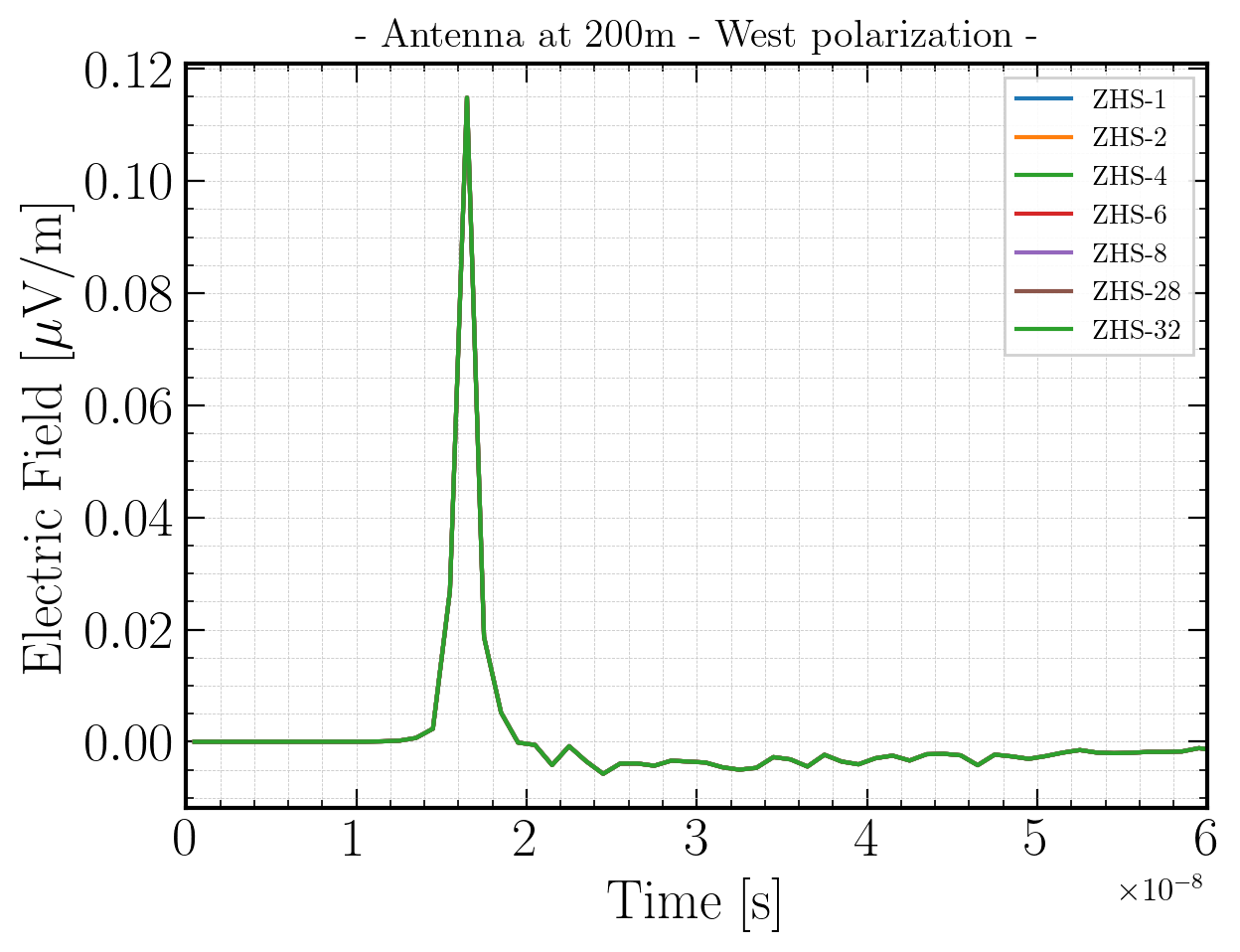}
    \label{fig:west-zhs}
\end{subfigure}
\begin{subfigure}{4.9cm}
    \includegraphics[width=4.9cm]{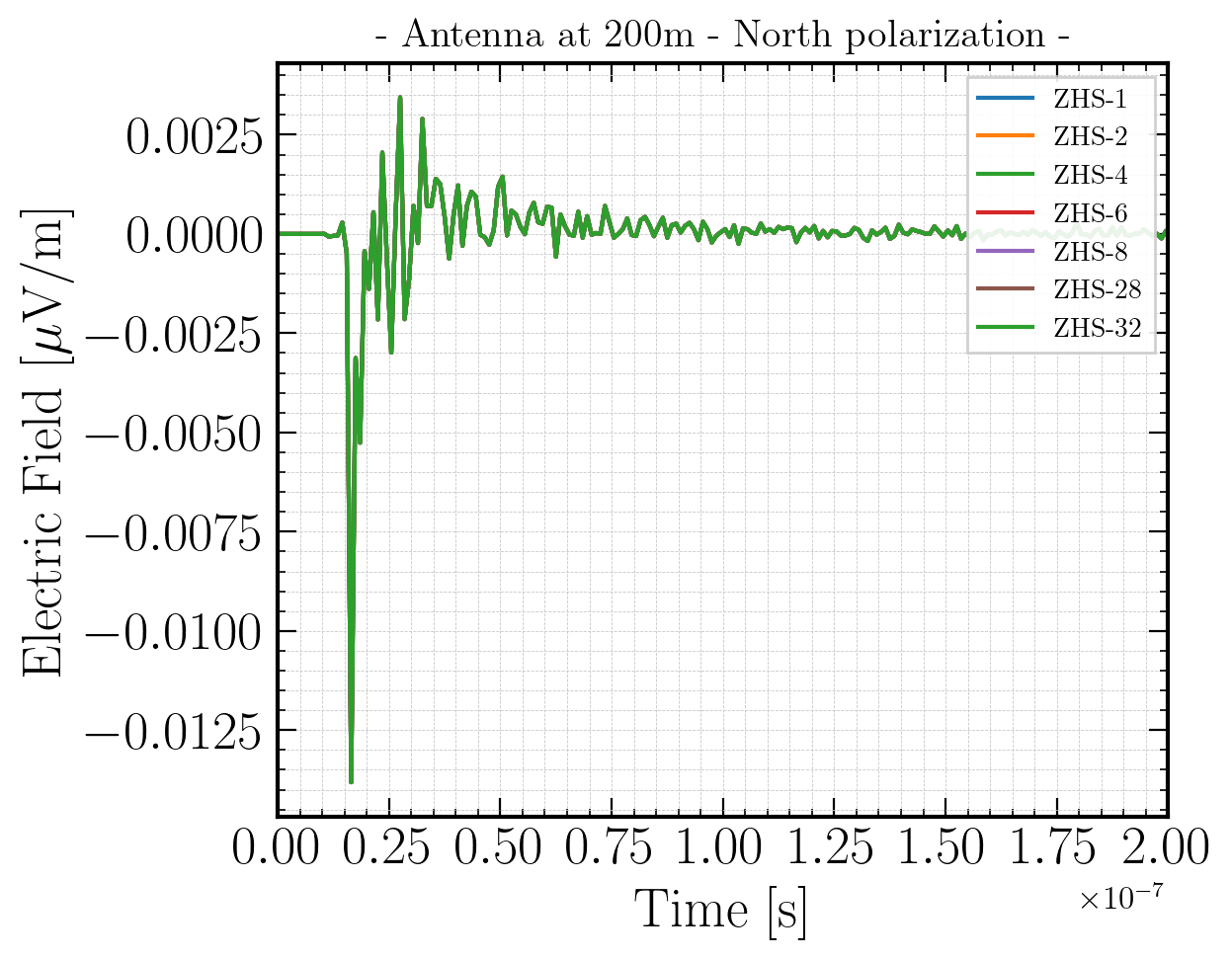}
    \label{fig:north-zhs}
\end{subfigure}
\begin{subfigure}{4.9cm}
    \includegraphics[width=4.9cm]{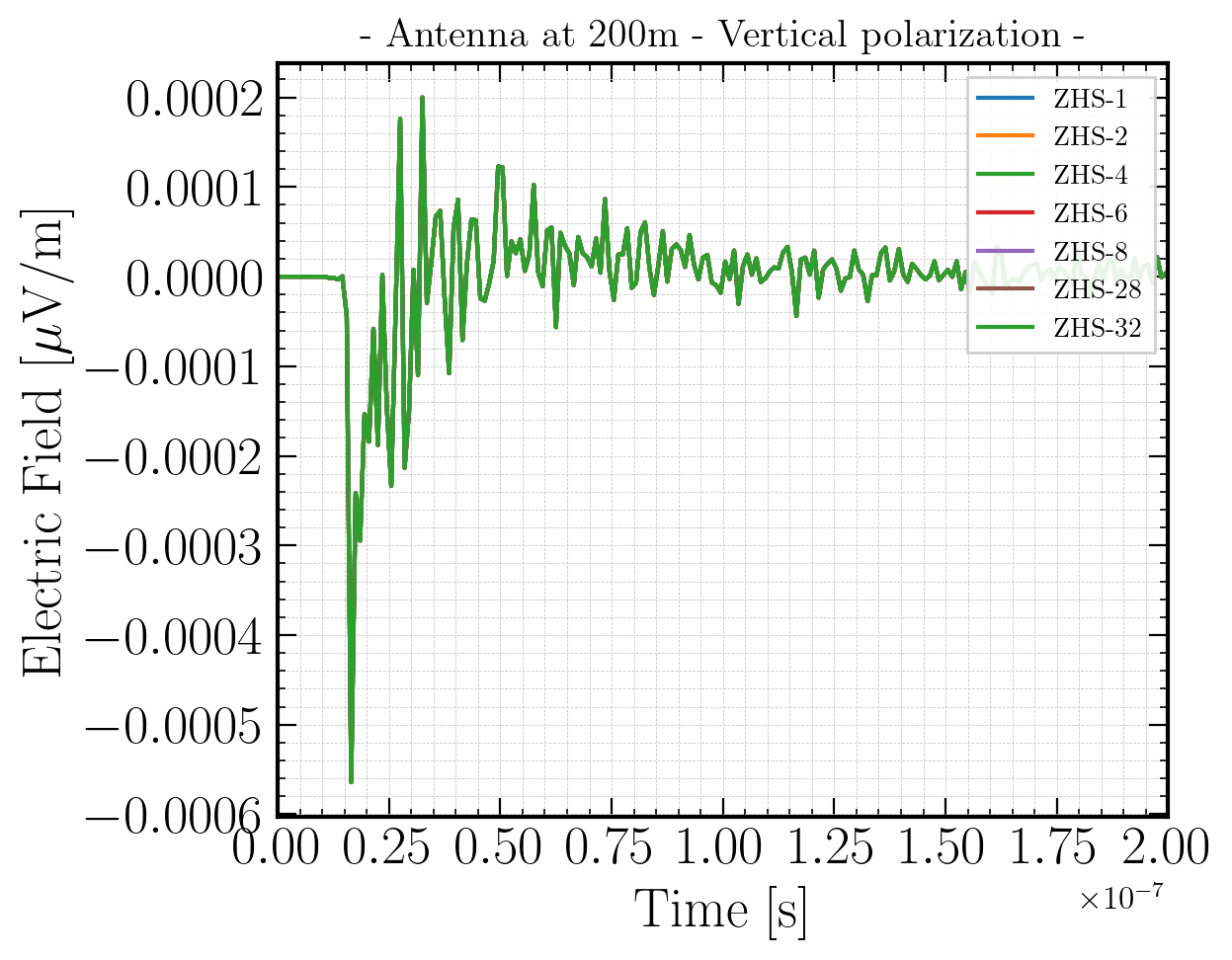}
    \label{fig:vertical-zhs}
\end{subfigure}
\caption{Pulse comparisons in all three polarizations using the ZHS formalism. The pulses have been simulated and processed on \cobj{gyges::gangs} of different sizes. Different number of threads produce identical pulses for ZHS, as expected.}
\label{fig:polarizations-zhs}
\end{figure}

Finally, the numerical consistence of the predictions for each algorithm has been checked for different numbers of threads. \autoref{fig:polarizations-coreas} and \autoref{fig:polarizations-zhs} show that there is no measurable impact of the parallelism in numerical results provided by each algorithm. The signal pulses simulated with both formalisms are identical regardless the number of threads, which confirms that the physics calculations are done consistently and accurately.

\section{Conclusions}
\label{sec:conclusions}

The status of the effort to parallelize the calculations of the radio module implemented in {CORSIKA 8} has been summarized. The implementation of the multithread dispatching mechanisms and management, which is based in Gyges, is compliant with C++17 or higher standard and allows specifying the number of worker threads without impacting any numerical result.
 The optimal number of threads, in which the performance peaks, depends on of the size of the antenna array. 
Under favorable, the performance gains are significant, with speeding-up reaching a factor 10 or superior. 
The code is currently under final internal review and should be integrated into CORSIKA 8 main branch in near future.   

\bibliography{PaperICRC2023.bib}

\newpage

\section*{The CORSIKA 8 Collaboration}
\small

\begin{sloppypar}\noindent
\input{latex_authorlist_authors}
\end{sloppypar}

\begin{center}
\rule{0.1\columnwidth}{0.5pt}
\raisebox{-0.4ex}{\scriptsize$\bullet$}
\rule{0.1\columnwidth}{0.5pt}
\end{center}

\vspace{-1ex}
\footnotesize
\input{latex_authorlist_institutions}

\vspace{-1ex}
\footnotesize
\input{acknowledgments}

\end{document}

%% file: latex_authorlist_authors.tex
J.M.~Alameddine$^{1}$,
J.~Albrecht$^{1}$,
J.~Alvarez-Mu\~niz$^{2}$,
J.~Ammerman-Yebra$^{2}$,
L.~Arrabito$^{3}$,
J.~Augscheller$^{4}$,
A.A.~Alves Jr.$^{4}$,
D.~Baack$^{1}$,
K.~Bernl\"ohr$^{5}$,
M.~Bleicher$^{6}$,
A.~Coleman$^{7}$,
H.~Dembinski$^{1}$,
D.~Els\"asser$^{1}$,
R.~Engel$^{4}$,
A.~Ferrari$^{4}$,
C.~Gaudu$^{8}$,
C.~Glaser$^{7}$,
D.~Heck$^{4}$,
F.~Hu$^{9}$,
T.~Huege$^{4,10}$,
K.H.~Kampert$^{8}$,
N.~Karastathis$^{4}$,
U.A.~Latif$^{11}$,
H.~Mei$^{12}$,
L.~Nellen$^{13}$,
T.~Pierog$^{4}$,
R.~Prechelt$^{14}$,
M.~Reininghaus$^{15}$,
W.~Rhode$^{1}$,
F.~Riehn$^{16,2}$,
M.~Sackel$^{1}$,
P.~Sala$^{17}$,
P.~Sampathkumar$^{4}$,
A.~Sandrock$^{8}$,
J.~Soedingrekso$^{1}$,
R.~Ulrich$^{4}$,
D.~Xu$^{12}$,
E.~Zas$^{2}$

%% file: latex_authorlist_institutions.tex
\begin{description}[labelsep=0.2em,align=right,labelwidth=0.7em,labelindent=0em,leftmargin=2em,noitemsep]
\item[$^{1}$] Technische Universit\"at Dortmund (TU), Department of Physics, Dortmund, Germany
\item[$^{2}$] Universidade de Santiago de Compostela, Instituto Galego de F\'\i{}sica de Altas Enerx\'\i{}as (IGFAE), Santiago de Compostela, Spain
\item[$^{3}$] Laboratoire Univers et Particules de Montpellier, Universit\'e de Montpellier, Montpellier, France
\item[$^{4}$] Karlsruhe Institute of Technology (KIT), Institute for Astroparticle Physics (IAP), Karlsruhe, Germany
\item[$^{5}$] Max Planck Institute for Nuclear Physics (MPIK), Heidelberg, Germany
\item[$^{6}$] Goethe-Universit\"at Frankfurt am Main, Institut f\"ur Theoretische Physik, Frankfurt am Main, Germany
\item[$^{7}$] Uppsala University, Department of Physics and Astronomy, Uppsala, Sweden
\item[$^{8}$] Bergische Universit\"at Wuppertal, Department of Physics, Wuppertal, Germany
\item[$^{9}$] Peking University (PKU), School of Physics, Beijing, China
\item[$^{10}$] Vrije Universiteit Brussel, Astrophysical Institute, Brussels, Belgium
\item[$^{11}$] Vrije Universiteit Brussel, Dienst ELEM, Inter-University Institute for High Energies (IIHE), Brussels, Belgium
\item[$^{12}$] Tsung-Dao Lee Institute (TDLI), Shanghai Jiao Tong University, Shanghai, China
\item[$^{13}$] Universidad Nacional Aut\'onoma de M\'exico (UNAM), Instituto de Ciencias Nucleares, M\'exico, D.F., M\'exico
\item[$^{14}$] University of Hawai'i at Manoa, Department of Physics and Astronomy, Honolulu, USA
\item[$^{15}$] Karlsruhe Institute of Technology (KIT), Institute of Experimental Particle Physics (ETP), Karlsruhe, Germany
\item[$^{16}$] Laborat\'orio de Instrumenta\c{c}\~ao e F\'\i{}sica Experimental de Part\'\i{}culas (LIP), Lisboa, Portugal
\item[$^{17}$] Fluka collaboration
\end{description}

%% file: acknowledgments.tex
\section*{Acknowledgments}
This research was funded by the Deutsche Forschungsgemeinschaft (DFG, German Research Foundation) – Projektnummer 445154105. For the simulations presented, computing resources from KIT have been used.